\documentclass[preprint,secnumarabic,amssymb, aps, prl,superscriptaddress]{revtex4}

\usepackage{graphicx}
\usepackage{amsfonts,amsmath,amssymb}

\begin{document}
\title{Domain wall tilting  in the presence of the Dzyaloshinskii-Moriya interaction in out-of-plane magnetized magnetic nanotracks}
\author{O. Boulle} 
\affiliation{SPINTEC, CEA/CNRS/UJF/INPG, INAC, 38054 Grenoble Cedex 9, France}
\author{S. Rohart} 
\affiliation{Lab. Physique des Solides, Univ. Paris-Sud, CNRS UMR 8502, 91405 Orsay, France}
 \author{L. D. Buda-Prejbeanu} 
\affiliation{SPINTEC, CEA/CNRS/UJF/INPG, INAC, 38054 Grenoble Cedex 9, France}
 \author{E. Ju\'{e}} 
\affiliation{SPINTEC, CEA/CNRS/UJF/INPG, INAC, 38054 Grenoble Cedex 9, France}
 \author{ I.M. Miron} 
\affiliation{SPINTEC, CEA/CNRS/UJF/INPG, INAC, 38054 Grenoble Cedex 9, France}
\author{ S. Pizzini} 
\affiliation{Institut N\'{e}el, CNRS, 25 avenue des Martyrs, B.P. 166, 38042 Grenoble Cedex 9, France}
\author{ J. Vogel} 
\affiliation{Institut N\'{e}el, CNRS, 25 avenue des Martyrs, B.P. 166, 38042 Grenoble Cedex 9, France}
\author{G. Gaudin}
\affiliation{SPINTEC, CEA/CNRS/UJF/INPG, INAC, 38054 Grenoble Cedex 9, France}
\author{A. Thiaville} 
\affiliation{Lab. Physique des Solides, Univ. Paris-Sud, CNRS UMR 8502, 91405 Orsay, France}

\date{\today}%
\begin{abstract}
We show that the Dzyaloshinskii-Moriya  interaction (DMI)  can lead to a  tilting of the domain wall (DW) surface in  perpendicularly magnetized magnetic nanotracks when  DW dynamics is driven by an easy axis magnetic field or a spin polarized current.   The DW tilting  affects the DW dynamics for large DMI and the tilting relaxation time can be very large  as it scales with the square of the track width. The results are well explained by an  analytical model based on a Lagrangian approach where the DMI and the DW tilting are included.  We propose a simple way to estimate the DMI in a magnetic multilayers by measuring the dependence of the  DW tilt angle on a transverse static magnetic field. Our results shed light on the current induced DW tilting observed recently in Co/Ni multilayers with inversion asymmetry, and further support the presence of DMI in these systems.
\end{abstract}

\maketitle
\

The effect of inversion asymmetry on the magnetic and electronic transport properties  at interfaces  of low dimensional magnetic film is currently attracting a growing attention.
In the presence of spin-orbit coupling, inversion asymmetry leads to an additional term in the exchange interaction, namely the Dzyaloshinskii-Moriya interaction (DMI)~\cite{Dzyaloshinskii57SPJ,Moriya60PR}, which tends to make the magnetization rotate around a local characteristic vector D.  This can destabilize the uniformly magnetized  states leading  to novel chiral magnetic order, such as   spin spiral~\cite{Bode07N,Tretiakov10PRL}. 
Novel out-of-equilibrium transport phenomena have also been demonstrated, such as current induced spin orbit torques induced by the Rashba spin orbit coupling and/or the spin Hall effect, leading to  current induced magnetization reversal~\cite{Miron11N,Liu12PRL,Liu12S}.
A recent striking example of the impact of the inversion asymmetry in ultrathin magnetic films is the current induced domain wall motion (CIDM) in perpendicularly magnetized nanotracks.  This was first outlined by Miron et al.  who reported very efficient CIDM in asymmetric Pt/Co(0.6~nm)/AlOx multilayers whereas   symmetric Pt/Co/Pt multilayers showed no effects~\cite{Miron09PRL,Miron11NMa}. The high perpendicular anisotropy in this material leads to very narrow DW ($\sim5$~nm), so that in typical experiments, the nanotrack width ($\sim100$~nm) is much larger than the DW width. It  is thus expected that the magnetization rotates parallel to the DW surface  (Bloch DWs) to minimize the magnetostatic energy. Whereas these experiments were first interpreted in terms of a high non-adiabatic torque induced by the Rashba spin orbit coupling, we recently proposed that the high efficiency arises from two key features resulting from  the inversion asymmetry  and the high spin orbit coupling in this material~\cite{Thiaville12EL}: First, a change of the DW equilibrium structure from Bloch to N\'eel induced by the  DMI.  This leads to chiral DWs where the DW magnetization rotates perpendicular to the DW surface with a unique sense of rotation~\cite{Heide08PRB,Chen13PRL};  Second, a large Slonczewski-like spin orbit torque  which is maximal  in the N\'eel configuration~\cite{Miron11N,Khvalkovskiy13PRB}. Recent  CIDM experimental results in Pt/Co/Ni~\cite{Ryu13NN} and Pt/CoFeB/MgO multilayers~\cite{Emori13NM}   seem to support this scheme. 

In this Letter, we show that   the inversion asymmetry not only affects the DW dynamics through a change of the internal DW structure but also through a modification of its shape. In perpendicular magnetized nanotracks, the DW surface is expected to be perpendicular to the nanotrack axis to minimize the DW length and thus its energy. However, in the presence of DMI, when driving the DW dynamics,  micromagnetic simulations reveal that a large DMI can lead to a sizable tilting of the DW surface  which can strongly affect  the DW dynamics. This DW tilting is a dynamical effect which occurs whatever the driving mechanism, e.g. an external magnetic field or a spin polarized current, and is thus  intrinsically different from the previously reported    current induced DW tilting~\cite{Partin74JAP,Viret05PRB, Yamanouchi06PRL}~\footnote{See  Supplementary Materials.}. The results are well explained using an  analytical model based on a Lagrangian approach where the DMI and the DW tilting are included. 
We also show  that the DW tilting can be controlled using a static transverse magnetic field, and this provides a simple way to measure directly the DMI experimentally. 
 Our results shed light on the unexplained current induced DW tilting recently observed in  Co/Ni  asymmetric multilayer nanotracks~\cite{Ryu12APE} and further support the presence of DMI in these materials~\cite{Ryu13NN}.

We consider a magnetic ultrathin film   grown on a substrate with a capping layer in a different material so that the inversion symmetry is broken along the vertical axis (z). The magnetization is supposed oriented out-of-plane with a strong perpendicular anisotropy. In addition to the standard micromagnetic energy density which includes the exchange, anisotropy, Zeeman and demagnetizing energy, we add the following DMI  that reads in a continuous form~\cite{Thiaville12EL} $E_{DM}=D\left[m_z\frac{\partial m_x}{\partial x}-m_x\frac{\partial m_z}{\partial x}+id.(x \rightarrow y)\right]$. This form corresponds to a sample isotropic in the plane, where the Dzyaloshinskii vector for any in-plane direction $\mathbf{u}$ is $D\mathbf{z}\times\mathbf{u}$ with $D$ a uniform constant, originating from the symmetry breaking at the z surface. 
Micromagnetic simulations are based on  the Landau-Lifschitz-Gilbert equation:
\begin{equation}\label{LLG}
\frac{\partial \textbf{m}}{\partial t}= -\frac{\gamma_0}{\mu_0M_s} \frac{\delta E}{\delta\textbf{m}}\times \textbf{m} + \alpha \textbf{m} \times \frac{\partial \textbf{m}}{\partial t} -\gamma_0 H_{SO}J\mathbf{m}\times(\mathbf{m}\times\mathbf{u_y})
\end{equation}
where $\gamma_0=\mu_0\gamma$ with $\gamma$   the gyromagnetic ratio, $E$ the energy density and $M_s$ the saturation magnetization. We assume that the injection of a current density $J$ in the nanotrack leads to  a  Slonczewski-like torque $-\gamma_0 H_{SO}J\mathbf{m}\times(\mathbf{m}\times\mathbf{u_y})$ ~\cite{Miron11N,Liu12S,Haney13PRB}.  To simplify, we do not consider the effect of the adiabatic and non-adiabatic spin transfer torque as well as the field like part of the spin orbit torque~\cite{Miron10NM}.  
	In the following, we consider  sufficiently large values of D ($D>0.12$~mJ/m$^2$ for our simulation parameters) so that the N\'{e}el configuration is stable at equilibrium~\cite{Thiaville12EL}.  2D micromagnetic simulations are performed using  modified homemade micromagnetic solvers~\cite{Szambolics09JMMM,Thiaville12EL}. The following parameters have been used~\cite{Miron11NMa}: exchange parameter A=$10^{-11}$ J/m,  saturation magnetization $M_s=1.09\times 10^6$ A/m,  uniaxial anisotropy constant $K=1.25\times10^6$ J/m$^3$,  Gilbert damping parameter $\alpha=0.5$,  thickness of $t_m=0.6$~nm.
	
\begin{figure}[p]
	\centering
		\includegraphics[width=1\textwidth]{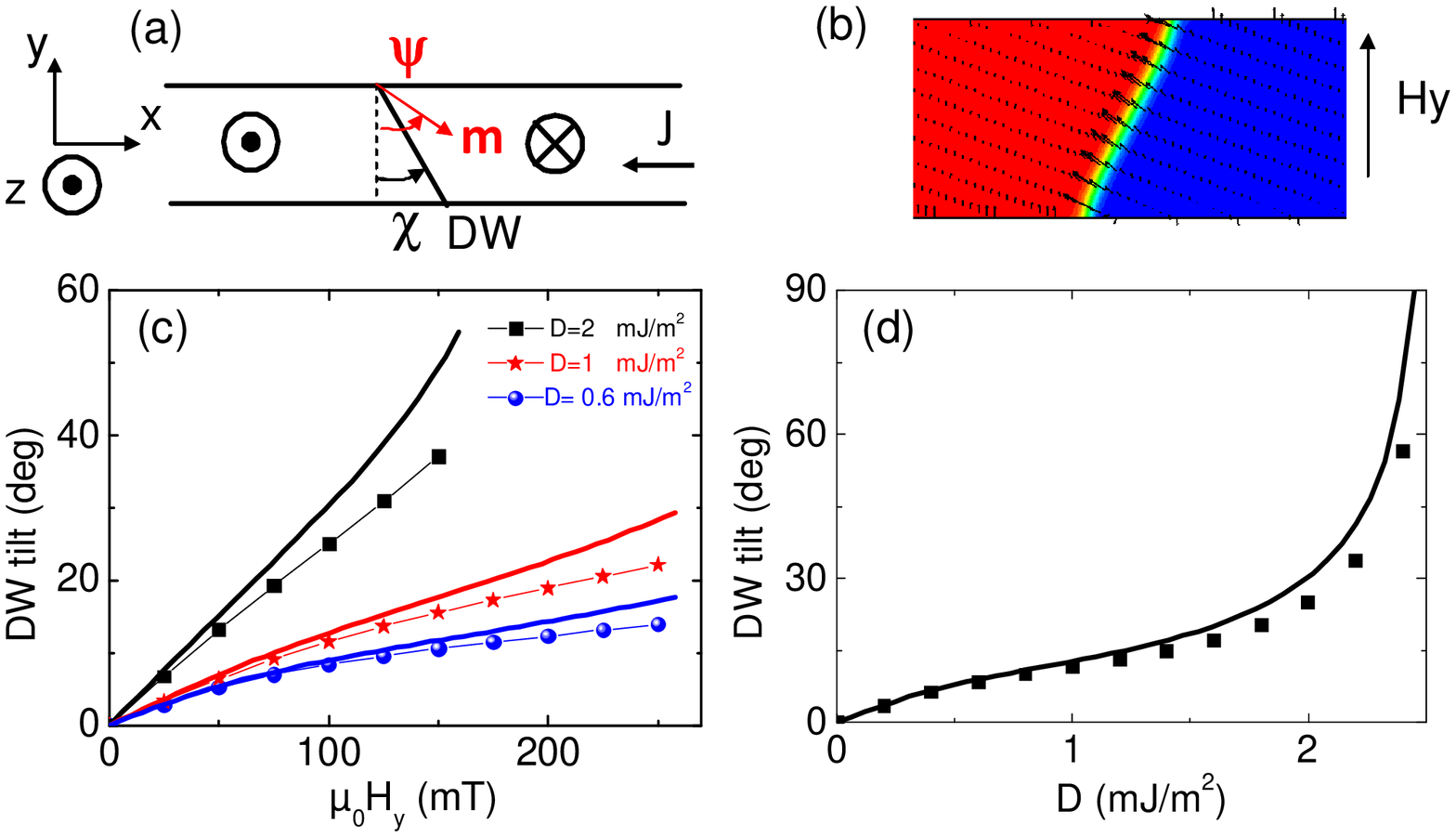}
	\caption{(a) Schematic of the tilted DW. 
	(b) Micromagnetic configuration of a 100 nm wide track with D=2 mJ/m$^2$ and a transverse magnetic field $\mu_0H_y=100$~mT. The color scale is  the same as Fig.~\ref{Fig2}(a).  (c) DW tilt angle  as a function of $\mu_0H_y$ for $D=2$ mJ/m$^2$ and (d)  as a function of  $D$ for $\mu_0H_y=100$~mT. Dots are the results of micromagnetic simulations whereas the continuous lines are the results of the CCM. 
} 
	\label{Fig1}
\end{figure}

The DW tilting induced by the DMI  can simply be introduced by considering the effect  of a static in-plane magnetic field $H_y$  transverse to the magnetic track (Fig~\ref{Fig1}(b)). In the presence of  $H_y$, the Zeeman interaction leads to a rotation of the DW magnetization away from the N\'{e}el configuration.  To   recover the N\'{e}el configuration  energetically favored by the DMI, the DW surface  tilts by an angle $\chi$ at the cost of a higher DW energy due to the larger DW surface.  Fig~\ref{Fig1}(b) shows the resulting DW tilting for $\mu_0H_y=100$ mT and a large value  $D=2$ mJ/m$^2$. The tilt angle as a function of $H_y$ and D  is plotted on Fig.~\ref{Fig1}(c,d). As expected, the DW tilting increases with $H_y$  and, for a fixed $H_y$, increases with D. 
The tilt angle can be roughly estimated from energetic considerations assuming that the DW always stays in a N\'{e}el configuration with an energy per surface unit $\sigma_0$. On the one hand, for a   DW tilted by an angle $\chi$,  the DW surface and thus the total energy is increased by a factor $1/\cos\chi$; on the other hand, the Zeeman energy per unit surface  scales as $\sigma_Z\sin\chi$ with $\sigma_Z=-\pi\mu_0H_yM_s\Delta$ ($\Delta$ is the DW width). This leads to a total DW energy  $E_{DW}\approx wt_m(\sigma_0-\sigma_Z\sin\chi)/\cos\chi$, where $w$ is the track width. The minimization of this energy leads to  $\sin\chi=\sigma_Z/\sigma_0$. 
  Importantly, the slope of the DW tilting as a function of $H_y$ on Fig.~\ref{Fig1}(c) depends directly on the value of $D$. This provides a direct way to measure $D$ experimentally, from the dependence of the  DW equilibrium tilt angle on $H_y$.

\begin{figure}[p]
	\centering
		\includegraphics[width=1\textwidth]{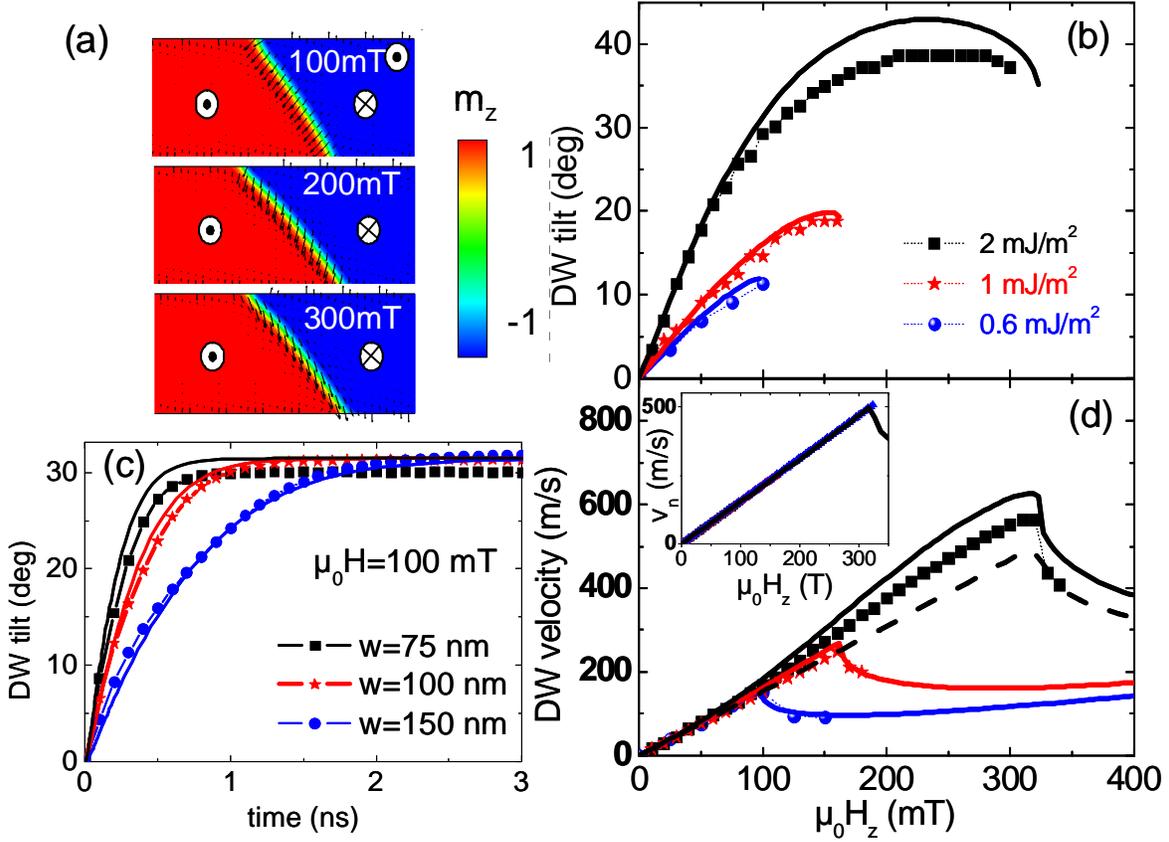}
	\caption{Dynamics of the DW driven by an external magnetic field $H_z$ for a 100 nm wide nanotrack.  (a) Magnetization pattern of the DW for different values of $H_z$ with D=2 mJ/m$^2$. Tilt angle (b)  and velocity (d) of the DW as a function of $H_z$ for different values of D. The inset in (d) shows the DW velocity $v_n$ in the direction perpendicular to the DW surface ($v_n=v\cos\chi$).  (c) Time dependence of the tilt angle for $\mu_0H_z=100$~mT and different track widths $w$ for $D=2$~mJ/m$^2$. In (b,c,d), the results of the micromagnetic simulation (resp. CCM) are plotted in coloured dots (resp. continuous lines). The dashed (resp. continuous) black line in (d) (resp (d), inset) corresponds to the prediction of the standard (q,$\psi$) model for $D= 2$ mJ/m$^2$.} 
	\label{Fig2}
\end{figure}

In the presence of DMI, a tilting of the DW surface can also be induced dynamically by applying an easy axis external magnetic field $H_z$. The magnetization distribution in the track for different magnetic fields   (Fig.~\ref{Fig2}(a),  $D=2$~mJ/m$^2$)   reveals that the DW tilts significantly when driven  by $H_z$ in the steady state regime.  As shown on Fig.~\ref{Fig2}(b), the steady-state tilt angle rapidly increases with $H_z$ and $D$, although a saturation is observed for large  $H_z$. Fig~\ref{Fig2}(d) shows the DW velocity $v$ along the track direction  as a function of  $H_z$ for different values of $D$. As expected, the DMI leads to an increase of the Walker field~\cite{Thiaville12EL}.  For large values of $H_z$, the DW velocity significantly deviates from the expected linear scaling as $D$ increases. This deviation is the direct result of the DW tilting: the propagation of the tilted DW at a velocity $v_n$ normal to its surface leads to a velocity $v=v_n/\cos{\chi}$ along the track direction. When considering $v_n$ instead of $v$ (Fig.~\ref{Fig2}(d), inset)), the expected linear scaling is recovered and the velocity in the steady state regime does not depend anymore on $D$.  The time dependence of the DW tilt angle is shown on Fig.~\ref{Fig2}(c) for several $w$ when applying $\mu_0H_z=100$~mT at t=0~\footnote{ See also the corresponding movie for $w=100$~nm in the Supplementary Materials.}. 


To describe the dynamics of tilted DWs induced by the DMI, we consider an extended collective coordinate model (CCM)~\cite{Slonczewski79} where the DW is described by three variables: its position $q$ in the track, the DW magnetization angle $\psi$ and the tilt angle  of the DW surface $\chi$ (cf Fig~\ref{Fig1}(a)).   The DW profile is described by the following Ansatz for the azimuthal  $\theta$ and polar angle $\varphi$: $\varphi(x,y,t)=\psi(t)-\pi/2$ and  $\theta=2\arctan[(x\cos\chi+y\sin\chi-q\cos\chi)/\Delta]$ ($\Delta=\sqrt{A/(K-\mu_0M_s^2/2)}$ ). 
The effect of the DMI on the DW profile and dynamics is taken into account by  an additional term in the DW energy (see below)~\cite{Thiaville12EL}. To derive the  dynamical equations, a Lagrangian approach is considered~\cite{Hubert74,Thiaville02JMMM,Boulle12JAP}.
 The LLG equation can be  derived by writing the Lagrange$-$Rayleigh equations for the Lagrangian $L=E+(M_s/\gamma)\varphi\dot{\theta}\sin\theta$ with $E$ the micromagnetic energy density  and $\textbf{m}=(\sin\theta\cos\varphi,\sin\theta\sin\varphi,\cos\theta)$. 
The effect of the damping and spin orbit torques is included in  the  dissipative function $F=\alpha M_s/(2\gamma)[d\mathbf{m}/dt-(\gamma_0/\alpha) H_{SO}J\mathbf{m}\times\mathbf{u_y}]^2$.

 The Lagrange-Rayleigh equations then leads to the following   CCM equations :
\begin{align}
\dot{\psi}+\frac{\alpha\cos\chi}{\Delta}\dot{q}= \gamma_0 H_z+\frac{\pi}{2}\gamma_0 H_{SO}J\sin\psi, \label{EQ:1Dmodel1} \\
\frac{\dot{q}\cos\chi}{\Delta}-\alpha\dot{\psi}= \frac{\gamma_0 H_k}{2}\sin2(\psi-\chi)+\frac{\pi D\gamma_0}{2\mu_0M_s\Delta}\cos(\psi-\chi)-\frac{\pi}{2}\gamma_0H_y\sin\psi,\label{EQ:1Dmodel2}\\
\dot{\chi}\frac{\alpha\mu_0 M_s\Delta\pi^2}{6\gamma_0}\left(\tan^2\chi+\left(\frac{w}{\pi\Delta}\right)^2\frac{1}{\cos^2\chi}\right)=\notag\\
-\sigma\tan\chi+\pi D \cos(\psi-\chi)+\mu_0H_kM_s\Delta\sin2(\psi-\chi)\label{EQ:1Dmodel3}
\end{align}
where $\sigma$ is the  wall energy per unit area with $\sigma=4\sqrt{AK}+\pi D\sin(\psi-\chi)+\mu_0H_kM_s\Delta\sin^2(\psi-\chi)+\pi\Delta M_sH_y\cos(\psi)$, with $H_k$ the DW demagnetizing field~\footnote{Note that the magnetization angle in the DW frame  $\Phi=\psi-\chi$ is the relevant one for the demagnetizing field  and DMI terms,  whereas the absolute angle $\psi$ is involved for the $H_{SO}$ and $H_y$ terms.}. 
 

Assuming that $\alpha w\gg \Delta$, these equations lead to a typical time scale for the tilting to settle $\tau=\alpha \mu_0M_s w^2/(6\sigma\gamma_0\Delta)$. The $w^2$ dependence is explained by the time to reverse the spins in the nanotrack surface swept by the DW when the tilting takes place. On the other hand, the magnetization angle in the DW frame relaxes on a shorter time scale $\tau_{\Phi}=\frac{1+\alpha^2}{\alpha\gamma}\frac{1}{\pi D/(2M_s\Delta)-H_k}$ which does not depend on $w$. 
In the steady state regime ($\dot{\chi}=0$, $\dot{\psi}=0$) and $H_y=0$, the tilt angle is directly related to the DW velocity $v$ as :
\begin{equation}
	\tan\chi=\frac{2M_s}{\gamma\sigma}v\cos\chi \label{EQ:tiltangle}
\end{equation}
with the DW velocity $v=\frac{\gamma_0 \Delta}{\alpha\cos\chi}(H_z+\frac{\pi}{2} H_{SO}J\sin\psi)$. This points to the dynamical origin of the DW tilting.   A more physical picture can be obtained from the expression of the  Lagrangian integrated  over   the nanotrack   $L_{DW}/(t_mw)=\frac{\sigma}{\cos\chi}-2\frac{M_s(\Phi+\chi)}{\gamma}\dot{q}$, where $\Phi$ is the magnetization angle in the DW frame ($\Phi=\psi-\chi$).   The first term is the DW internal energy proportional to the DW surface and thus scaling as $1/\cos\chi$. The second term can be seen as a kinetic potential~\cite{Hubert74}, which contrary to a kinetic energy, is linear with the DW velocity and the DW angle.  For the field driven case in the steady state regime, $\Phi$ is  defined only by the in-plane torques   due to $H_z$, $D$ and $H_k$ (see Eq.~\ref{EQ:1Dmodel1}-\ref{EQ:1Dmodel2})  and does not depend on $\chi$. The tilt angle in the steady state regime  can thus be deduced   from the minimization of $L_{DW}$ with $\chi$ at fixed $\Phi$, which leads to Eq.~\ref{EQ:tiltangle}~\footnote{Note that this expression is  general. Very small DW tilts  are also predicted in the absence of DMI, but the DMI leads to large tilt angle as it allows   fast steady state DW motion and it decreases the internal DW energy.}. The tilt angle is thus the result  of a balance between the gain in the kinetic potential resulting from the DW tilting and the cost in the increased DW energy due to the larger surface. 


We now compare the predictions of this model with the results of the  micromagnetic simulations.  The continuous lines on Fig~\ref{Fig1}(c,d) show the DW tilting induced by  $H_y$ predicted by the CCM whereas the DW tilt angle, time dependence  and DW velocity driven by $H_z$ are plotted in continuous  lines on Fig.~\ref{Fig2}(b-d) : a general good agreement  is obtained with the micromagnetic simulations  despite the simplicity of the model. 
 We also plotted the results  of the standard  ($q$,$\psi$) model on Fig~\ref{Fig2}(d) (dashed line). The model  does not reproduce the nonlinear increase of the DW velocity, but a good agreement is obtained when considering  the DW velocity in the direction perpendicular to its surface $v_n$ (inset).  The DW tilting thus does not affect the DW velocity perpendicular to its surface. 

We now consider the current driven DW dynamics induced by the Slonczewski-like spin orbit torque   in the presence of a large DMI. This torque is expected for structure with inversion asymmetry such as Pt/Co/AlOx trilayers~\cite{Miron10NM,Miron11N,Miron11NMa}. It may arise from the spin Hall effect due to the current flowing in the non magnetic layer and/or from the Rashba spin orbit interaction~\cite{Miron11N,Liu12S}. 
It leads to an effective easy-axis magnetic field on the DW $H_{SO}J$ proportional to $\sin\psi$ (see Eq.~\ref{EQ:1Dmodel1}), which is thus maximal for a N\'{e}el DW configuration ($\psi=\pm\pi/2$) obtained for sufficiently high $D$. The field $H_{SO}$ can be very large $\sim0.07$ T/(10$^{12}$ A/m$^2$) in  Pt/Co/AlOx~\cite{Garello13NN} (see also Ref.~\cite{Pi10APL, Kim12NM,Liu12PRL,Emori13NM}).    Similarly to the action of  $H_y$, the spin orbit torque tends to rotate the DW magnetization  along the y direction away from the N\'eel configuration, providing an additional source for the DW tilting. 

\begin{figure}[p]
	\centering
		\includegraphics[width=1\textwidth]{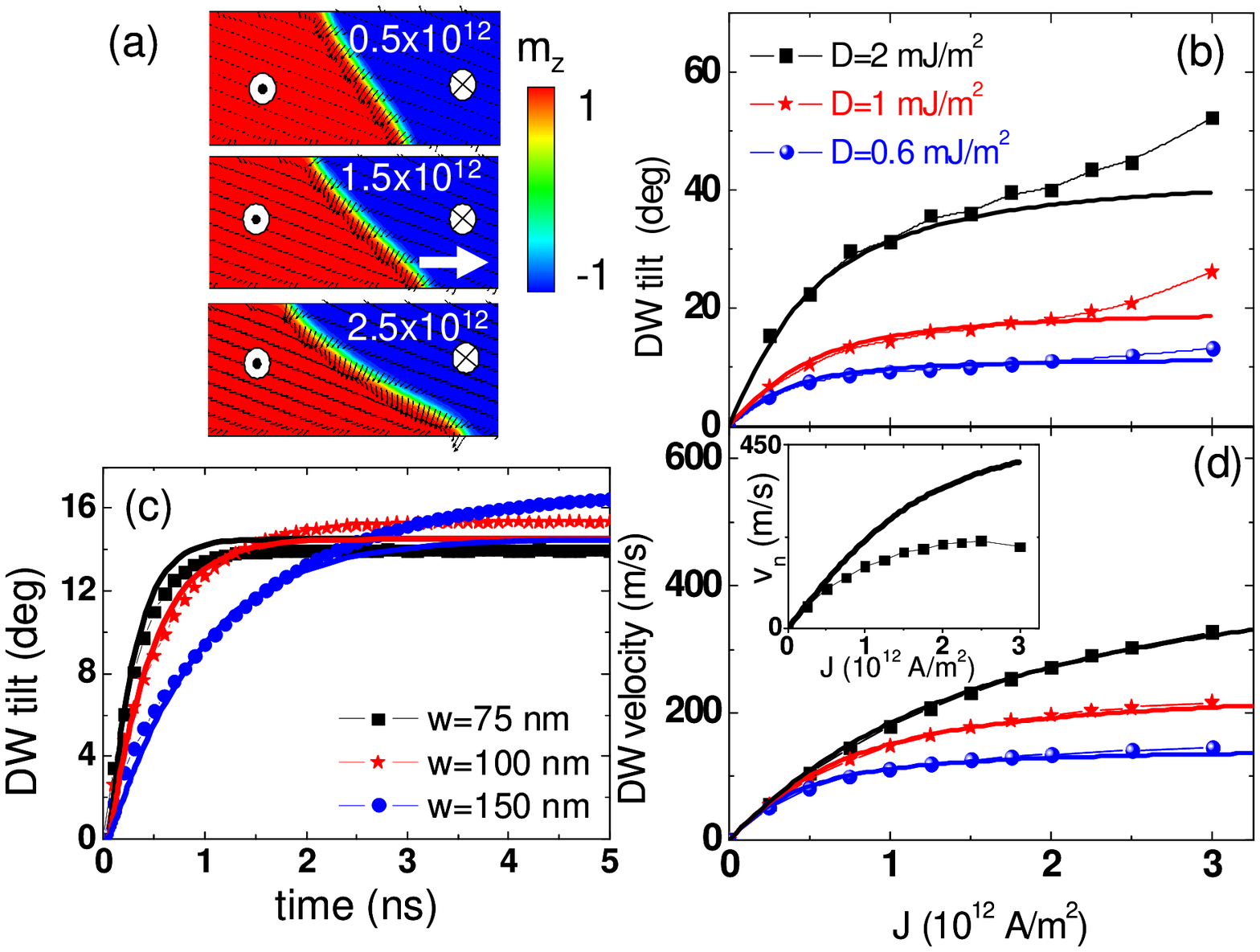}
	\caption{Dynamics of the DW driven by the spin orbit torque  ($\mu_0H_{SO}=0.1$~T/($10^{12}$~A/m$^2)$) for a 100 nm wide nanotrack. The results of the micromagnetic simulation (resp. CCM) are plotted in coloured dots (resp. continuous lines). (a) Magnetization pattern of the  DW for  $D=2$~mJ/m$^2$ and different values of $J$. The white arrow indicates the current direction.  (b,d)  Tilting (b) and velocity (d) of the DW as a function of $J$ for different values of $D$. The inset in (d) shows the DW velocity in the direction perpendicular to the DW surface $v_n$  for $D=2$ mJ/m$^2$.   The continous black line is the result of the standard ($q$,$\psi$) model. (c)  Time dependence of the DW tilt angle for different track widths $w$  for a current  of density $0.25\times10^{12}$~A/m$^2$ applied at t=0  (D=2 mJ/m$^2$). }
	\label{Fig3}
\end{figure}

 The results of micromagnetic simulations of the DW dynamics driven by the spin orbit torque with $\mu_0H_{SO}=0.1$ T/(10$^{12}$ A/m$^2$) are shown on Fig.~\ref{Fig3}. When injecting a current in the track, a fast DW motion  is observed  against the electron flow and the velocity increases with $J$ and $D$ (see Fig.~\ref{Fig3}(d)). At the same time, a significant tilting of the DW occurs (see Fig.~\ref{Fig3}(a) for $D=2$~mJ/m$^2$), which increases with $J$ and $D$ (Fig.~\ref{Fig3}(b)). The DW velocity and the tilting predicted by the CCM are shown on Fig.~\ref{Fig3}(b,d), continous lines. An excellent agreement is obtained with the micromagnetic simulations except at higher current densities for the tilt angle  due to the onset of a more complex DW structure (see Fig.~\ref{Fig3}(a) for $J=2.5\times10^{12}$~A/m$^2$). The  DW velocity in the direction perpendicular to the DW surface $v_n$ for $D=2$~mJ/m$^2$ is plotted Fig.~\ref{Fig3}(d), inset. 
Contrary to the field driven case, the standard ($q$,$\psi$) model strongly overestimates the DW velocity  (continous line). As expected, the  DW tilting leads to an additional rotation of the DW angle $\psi$ away from $\pi/2$ where the torque is maximal. The DW tilting thus leads to a large decrease of the DW velocity. This clearly illustrates the importance of the DW tilting on the CIDWM for large DMIs. Fig.~\ref{Fig3}(c) shows the time dependence of the tilting for a current pulse of $J=0.25\times10^{12}$~A/m$^2$ applied at t=0. The CCM (continuous lines)  reproduces well the time scale for the tilting to take place which scales as $w^2$. 



   

Experimentally,  Ryu~\textit{et al.} recently reported fast current induced DW motion associated with a significant DW tilting in asymmetric perpendicularly magnetized (Pt/Co/Ni/Co/TaN)   nanotracks~\cite{Ryu12APE,Ryu13NN}.  By studying the dependence of the current induced DW velocity on an in-plane longitudinal magnetic field, they present evidence of  chiral DWs driven by the Slonczewski like spin-orbit torque in agreement with the presence of a DMI. The DW tilting is reversed for up/down and down/up  DW which is well explained by N\'{e}el DWs pointing in opposite directions due to the DMI. 
From the longitudinal magnetic field required to suppress the CIDM  and using the magnetic and transport parameters of Ref.~\cite{Ryu12APE,Ryu13NN}, one can deduce a DMI of $D=0.8$~mJ/m$^2$  for A=$1\times10^{-11}$~J/m. Using this value, micromagnetic simulations predict  a steady state tilt angle of about $18^\circ$  for $J=1\times10^{12}$~A/m$^2$ close to the one measured experimentally ($\sim20^\circ$)~\footnote{The following parameters have been used :  $M_s=0.6\times10^6$~A/m, $K_u=0.59\times10^6$ J/m$^3$,   $\mu_0H_{SO}=4.8\times10^{-14}$~T/(A/m$^2$) corresponding to a spin Hall angle of 0.1, a damping constant $\alpha=0.05$, a spin polarization $P=0$.}.  Smaller additional contributions   may also arise from the anomalous Hall effect and the Oersted field~\footnote{See  Supplementary Materials.}. Our model thus accounts for the  DW tilting reported by Ryu \emph{et al.} which further supports the presence of DMI in these inversion asymmetric multilayers. 

To conclude,  we have shown that the DMI can lead to a  tilting of the DW surface in  perpendicularly magnetized  nanotracks when   DW dynamics is driven by an easy axis magnetic field or a spin polarized current. The DW tilting is the result of the balance between the gain in the kinetic potential of the moving DW when tilted and the increased DW energy due to  the larger DW surface. 
The DW tilting  affects the DW dynamics for large DMI and the tilting relaxation time can be very large as it scales with the square of the track width. We propose a simple way to estimate the DMI in magnetic multilayers by measuring the dependence of the  DW tilt angle on a transverse static magnetic field. Our results shed light on the current induced DW tilting observed recently in perpendicularly magnetized Co/Ni multilayers with inversion asymmetry and further support the presence of DMI in these materials.
This work was supported by  project Agence Nationale de la Recherche, project   ANR 11 BS10 008 ESPERADO.

\appendix
\section{ Supplementary Materials: Alternative mechanisms for current induced domain wall tilting in magnetic nanotracks}
In this section, we discuss other mechanisms that may lead to current induced DW  tilting  in  magnetic nanotracks.



\subsection{Anomalous Hall effect}
When  injecting a current in a perpendicularly magnetized nanotrack, the Anomalous Hall effect (AHE) leads to charge accumulations on the edges of the track  whose sign depends on the out-of-plane orientation of the magnetization. Around a magnetic domain wall (DW), the abrupt change in the charge polarity  leads to a deviation of the current lines around the DW which results in an additional current density $+\Delta J$ (resp. $-\Delta J$) on the right (resp. left)  edge of the track. This effect was first described by   Partin et al.~\cite{Partin74JAP} in 1974 and is at the basis of the DW drag effect introduced by Berger~\cite{Berger74JPCS}: The current loop   around the DW induced by the AHE creates an additional out-of-plane magnetic field which tends to move the DW in the direction of the charge carrier. This effect is mostly prevalent in narrow and thick nanotracks. For nm thin  magnetic nanotracks, the  DW drag effect is generally very small compared to the spin transfer/spin orbit torque~\cite{Viret05PRB}. However,  the additional current densities $\pm\Delta J$ on the edges   translate into different DW velocities $v=v_0+\pm\Delta v$    which can create   DW tilting. Assuming  a constant velocity   with time,   the tilting angle $\chi$ increases with the pulse length $\tau$ as:
\begin{equation}
\tan\chi=\frac{2\tau }{w}\Delta v\approx\frac{2\tau}{w}\left(\frac{dv}{dJ}\right)_J\Delta J
	\label{Eq:DWtilting}
\end{equation}

where $w$ is the track width. Partin~et al.~\cite{Partin74JAP} calculated the current density distribution  in the nanotrack  for a non-tilted DW. To first order in the Hall angle, the additional current density $\Delta J$ at the right and left  edge of the track reads :
\begin{equation}
	\frac{\Delta J}{J}=\pm\frac{40\tan\theta_H}{9\pi}
\end{equation}
 where J is the injected current density.
 
For a   Hall angle of  $1\%$ typically observed in perpendicularly magnetized ultrathin multilayers,  this leads to an additional current density of about $\pm1.5\%$ at the right and left edge of the track.
Ryu \emph{et al.} reported in Ref.~\cite{Ryu12APE} current induced DW tilting in  Pt(1.5)/Co(0.3)/Ni(0.7)/Co(0.15)/TaN(5) (thickness in nm). Assuming an Hall angle $<2\%$, one obtains $\Delta J/J<3\%$. For a DW velocity of 100 m/s for a current density of  $1\times10^{12}$ A/m$^2$, Eq.~\ref{Eq:DWtilting} leads to a DW tilting angle $<3^\circ$ for the 100 ns pulses used in their experiments.  This is about an order of magnitude lower than the one observed experimentally ($\sim20^\circ$). Reciprocally, the DW tilting observed experimentally would need an unrealistically large Hall angle of about $16\%$. Thus the anomalous Hall effect cannot account for the reported DW tilting but may add a small   contribution.

Current induced DW tilting was also reported by Yamanouchi \emph{et al. }in (Ga,Mn)As diluted magnetic semiconductor nanotracks~\cite{Yamanouchi06PRL,Yamanouchi07S}. The nanotracks were typically 5~$\mu$m wide and 30~nm thick. In Ref.~\cite{Yamanouchi06PRL}, the authors report   a large DW tilting (up to 60$^\circ$) for  $J<1\times10^{10}$ A/m$^2$ which increases with the pulse width. 
In these experiments, the large thickness of the magnetic films makes unlikely the presence of Dzyaloshinskii-Moriya (DMI). However, (Ga,Mn)As may have very large Hall angles (5 to 10 $\%$~\cite{Edmonds03JAP,Yamanouchi06PRL})  and  the authors used  long current pulses (up to 20 $\mu$s).  Assuming an Hall angle of 0.1, the additional  current  circulating around the  DW due to the anomalous Hall effect has a density $\Delta J/J=15\%$. Using Eq.~\ref{Eq:DWtilting} and for a DW velocity of about  1.75 m/s for $J=4\times10^9$ A/m$^2$, one can estimate   a DW tilting of about 60$^\circ$ in agreement with the experimental results. In the case of (Ga,Mn)As, the anomalous Hall effect seems thus to be the main driving force which creates the DW tilting~\footnote{The Oersted field may also contribute to the DW tilting but the resulting DW tilting is expected to be much lower (a few degree) due to the low injected current density~\cite{Yoo13JMMM,Yamanouchi06PRL}.}.

\subsection{Oersted  field effect }

The Oersted field created by the current circulating in the nanotrack has a perpendicular component which is antisymmetric with respect to the center of the track and can reach relatively large values on the edges (up to 5.2 mT at $J=1\times 10^{12}$ A/m$^2$  for the 10~$\mu$m wide and 2.65~nm thick   track     used in the experiments of Ryu~\emph{et al.}~\cite{Ryu12APE,Yamanouchi06PRL,Yoo13JMMM}). It  may thus favor a tilting of the DW. However, in the experiments of Ryu~\emph{et al.}~\cite{Ryu12APE},  the Oersted field  should tilt the DW in a direction opposite  to what is observed experimentally  and thus  can not   explained the reported DW tilting. It may however decrease the steady state DW angle  as it goes against the effect of the DMI.
 
  To study the effect of the Oersted field on the DW tilting, we carried out micromagnetic simulations taking into account the Oersted field for a 300~nm wide and 50 $\mu$m long track.  The following parameters have been used to mimic the experiments of Ryu~\emph{et al.}~\cite{Ryu12APE,Ryu13NN} : $A=1\times10^{11}$ J/m, $D=0.8$~mJ/m$^2$, $M_s=0.6\times10^6$~A/m, $K_u=0.59\times10^6$ J/m$^3$,  a spin orbit torque effective field $\mu_0 H_{SO}=4.8\times10^{-14}$~T/(A/m$^2$) corresponding to a spin Hall angle of 0.1, a damping constant $\alpha=0.05$, a spin polarization $P=0$, a buffer layer of Pt of 1.5~nm and a Co/Ni  thickness of 1.15~nm. The current density is assumed homogeneous in the track.  In the absence of Oersted field, the steady state angle is 18.3$^\circ$ for $J=1\times10^{12}$~A/m$^2$, close to the predictions of the collective coordinate model (17.3$^\circ$). For this wire width, the Oersted field leads only to a slight decrease of the steady state angle (17.6$^\circ$). For a given current density,  larger effects of the Oersted field are expected for larger width~\cite{Yoo13JMMM}. To study the effect of the track width, we carried out  micromagnetic simulations for a 2~$\mu$m wide    nanotrack ($J=1\times10^{12}$~A/m$^2$). Due to the much larger numbers of cells and the larger DW tilting relaxation time, the steady state was not fully reach in this case. However, one can estimate that the  Oersted field leads to a decrease of the DW tilting by about  $14-15\%$, which leads to a steady angle of   $15-16^\circ$. Experimentally, Ryu~\emph{et al.} reported tilt angles up to    $\sim13^\circ$ in 2~$\mu$m wide wires ($J=1.1\times10^{12}$~A/m$^2$).

For larger width ($w=5$, 10 and 20 $\mu$m),   tilt angles of around 20$^\circ$   ($J=1.1\times10^{12}$~A/m$^2$) have been reported  by Ryu~\emph{et al.} and the tilt angle   depends  little on the track width.  Had the Oersted a large impact on the tilt angle, a large decrease of the tilt angle as $w$ increases would be expected~\cite{Yoo13JMMM}, which is not observed. We can thus conclude that the Oersted field has little impact on the steady state tilt angle in  these experiments.


\subsection{Domain wall deformation in soft in-plane magnetized nanotracks}
The    DW deformation  of  moving DWs in soft in-plane magnetized nanotracks is a well established phenomenon~\cite{Nakatani03NM,Thiaville06,Lee07PRB,Klaui05PRL,Heyne08PRL,Klaui08JPCMa,Clarke08PRB}.   In the absence of    magnetocrystalline anisotropy, the equilibrium internal DW structure in soft nanotrack is defined by the demagnetizing and the exchange energy and thus depends only on the track aspect ratio for a given material. This leads to wide DWs whose internal structure  can be  easily deformed as compared to DWs in materials with strong magnetocrystalline anisotropy.  Two types of DWs are typically found in such tracks, head-to-head vortex DWs (VDW) and transverse DWs (tDW)~\cite{Klaui08JPCMa}.  When driving a DW dynamics by an external magnetic field or a spin polarized current, complex DW deformations have been predicted~\cite{Nakatani03NM,Thiaville06,Lee07PRB,Clarke08PRB}. For example, in the viscous regime, the VDW dynamics is characterized by  a shift of the transverse position of the vortex core, whereas in the precessional regime, periodic  DW transformations from tDWs to vortex/antivortex DWs occur~\cite{Nakatani03NM,Lee07PRB,Clarke08PRB}. Such periodic transformations of tDW into VDW  were observed experimentally in permalloy nanotracks, induced by the injection of consecutive current pulses~\cite{Heyne08PRL}.  However,  these changes  are in fact  a modification of the internal DW structure induced by the precessional torque and not a tilting of the DW surface.  For a standard Bloch DW, this is physically  equivalent to the in-plane rotation of the DW magnetization. In terms of our collective coordinate model, the DW deformations in VDW/tDW is equivalent to a change of the variable $\psi$ which describes the DW magnetization (see Ref.~\cite{Thiaville07EPJB,Clarke08PRB} for a more detailed discussion) but not a change of the inclination of the DW surface $\chi$. A tilting of the DW  surface in soft magnetic nanotracks were predicted in the special case of a DW moving in the presence of a transverse in-plane magnetic field~\cite{Bryan08JAP}, which may relate closer to the present study  as the DMI acts similarly to a constant external magnetic field exerted on the DW.    

  	\end{document}